\documentclass[%
 reprint,
 amsmath,amssymb,
 aps,
floatfix,
]{revtex4-1}
\usepackage{epsf}
\usepackage{graphicx}
\usepackage{dcolumn}
\usepackage{bm}

\begin{document}


\title{Experimental evidence for Tayler instability in a 
liquid metal column}

\author{Martin Seilmayer}
\author{Frank Stefani}%
\email{F.Stefani@hzdr.de}
\author{Thomas Gundrum}
\author{Tom Weier}
\author{Gunter Gerbeth}

\affiliation{Helmholtz-Zentrum Dresden-Rossendorf, P.O. Box 510119, D-01314 Dresden, Germany}%

\author{Marcus Gellert}
\author{G\"unther R\"udiger}
\affiliation{Leibniz-Institut f\"ur Astrophysik Potsdam, 
An der Sternwarte 16, D-14482 Potsdam, Germany}%

\date{\today}

\begin{abstract}
In the current-driven, kink-type Tayler instability (TI) a 
sufficiently strong 
azimuthal magnetic field becomes unstable against 
non-axisymmetric perturbations. 
The TI has been discussed 
as a possible ingredient of the solar dynamo mechanism and a source of the helical 
structures in cosmic jets. It is also considered as a size limiting 
factor for liquid metal batteries. We report 
on a liquid metal TI experiment using a cylindrical column of the 
eutectic alloy GaInSn to which electrical currents of up to 8 kA are applied. 
We present results of external magnetic field measurements that 
indicate the occurrence of the TI in good agreement with numerical 
predictions. The interference of TI with the competing 
large scale convection, resulting from Joule heating, is also discussed.
\end{abstract}

\maketitle


The last years have seen  a number of liquid metal experiments 
on various magnetohydrodynamic instabilities 
with relevance to the origin and the action of cosmic magnetic fields 
\cite{ZAMM}. Dynamo action has been observed in the large scale 
liquid sodium experiments in Riga, Karlsruhe, 
and Cadarache \cite{DYNAMO}. The helical version 
of the magnetorotational instability (MRI) has been evidenced
in the PROMISE experiment \cite{PROMISE}, and  further experiments in 
Maryland \cite{SISAN} and Princeton \cite{PRINCETON} 
are devoted to the investigation of 
the standard version of MRI. What is missing yet in the 
liquid metal lab is any evidence of the Tayler instability (TI) 
\cite{VANDAKUROVTAYLER}.
This is in remarkable contrast to the vast experience in plasma physics 
where the (compressible) counterpart of 
TI is better known as the kink instability in a z-pinch \cite{BERGERSON}, i.e. 
the limit of the Kruskal-Shafranov
instability when the safety factor goes to zero. 
In astrophysics, TI has been discussed as 
a possible ingredient of an alternative, nonlinear stellar dynamo 
mechanism (Tayler-Spruit dynamo \cite{SPRUIT}), as a
generation mechanism for helicity \cite{Gellert_1}, 
and as a possible source of helical structures
in galactic jets and outflows \cite{MOLL}. 

\begin{figure}[h]
\begin{center}
\epsfxsize=8.5cm\epsfbox{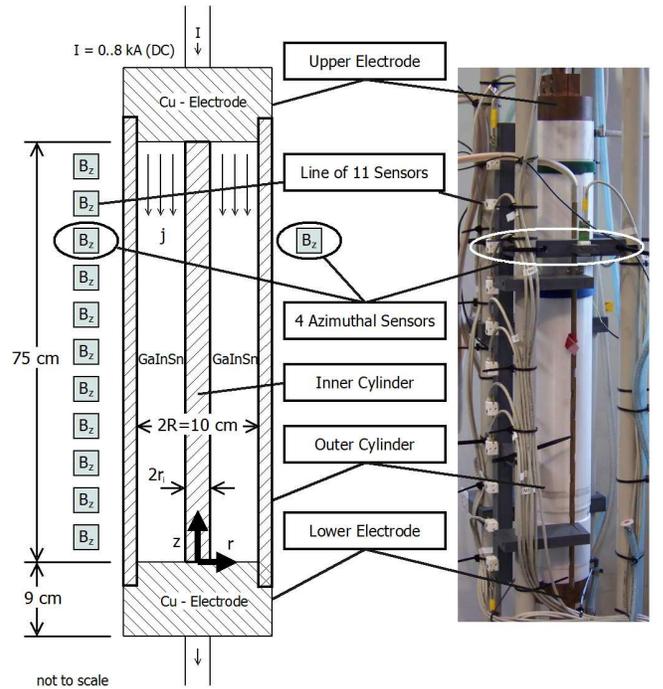}\\
\vspace{2mm}
\caption{Experimental set-up. Left: Scheme 
with liquid metal column and fluxgate sensors
positioned along the vertical axis and the azimuth. 
Right: Photograph of the central part of the experiment.}
\end{center}
\end{figure}

A particular motivation to study the TI in a liquid metal arises
from the growing interest in large-scale liquid metal batteries
as cheap means for the storage of highly intermittent 
renewable energies. 
Such a battery would consist of a self-assembling stratification of a 
heavy liquid half-metal (e.g. Bi, Sb) at the bottom, an 
appropriate molten salt as electrolyte in the middle, 
and a light alkaline or earth alkaline metal (e.g. Na, Mg) at the top. 
While small versions of this battery have already been tested 
\cite{BRADWELL}, 
for larger versions the occurrence of TI could represent a 
serious problem for the integrity of the stratification. 
In a recent paper \cite{CONVERSION} 
we have proposed a simple trick to avoid the TI in 
liquid metal batteries by just returning the battery current 
through a bore in the middle. By the resulting change of the radial 
dependence of $B_{\varphi}(r)$ it is possible to prevent the 
condition for (ideal) TI,
$\partial (r B^2_{\varphi}(r))/\partial r>0$.
Without such a provision, TI is expected to set in at
some finite electrical current in the order of kA.
The precise value is a fucntion of  
various material parameters,
since for viscous and resistive fluids TI is 
known \cite{RUEDIGER2} 
to depend effectively on the Hartmann number 
$Ha=B_{\varphi} R (\sigma/(\rho \nu)^{1/2}$, where $R$ is 
the radius of the column,  
$\sigma$ the electrical conductivity, 
$\rho$ the density, and 
$\nu$ the kinematic viscosity of the fluid. 

In this paper, we present experimental results that confirm 
the numerically
determined growth rates of TI \cite{RUEDIGER2} as well as the
corresponding prediction  that the critical
current increases monotonically with the 
radius of an inner cylinder 
\cite{CONVERSION,RUEDIGER3}. 

\begin{figure}[h]
\begin{center}
\epsfxsize=8.2cm\epsfbox{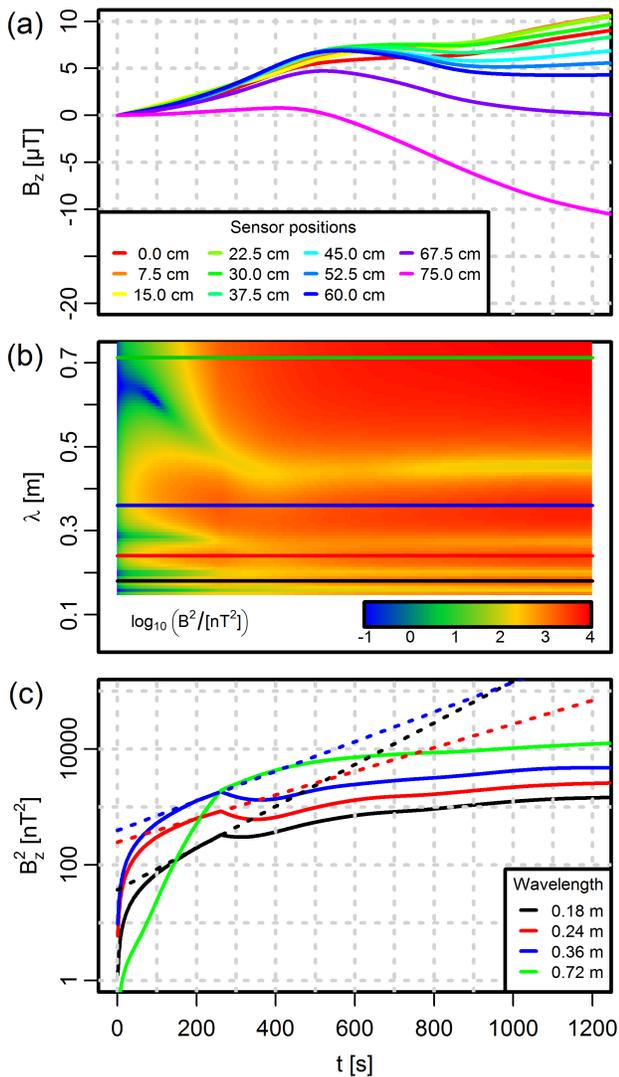}\\
\vspace{2mm}
\caption{Magnetic fields measured along the vertical axis, for
$r_i$=12.5 mm and $I$=7 kA.
(a) Time dependence of $B_z$ at the 11 fluxgate sensors.  (b) 
Power spectral density in dependence on time and wavelength. 
Note the 
dominance of short wavelength signals in the initial phase, 
and the dominance of the
long wavelength signal at later times. 
(c) Detailed time evolution of the PSD for
four significant wavelengths, showing a transition to an 
exponential growth around $t\sim$150 s, which is later
stopped at $t \sim$280\,s.}
\end{center}
\end{figure}

The central part of our TI experiment (Fig.~1) is an insulating cylinder 
of length 75\,cm and inner diameter 10\,cm, filled with the eutectic 
alloy GaInSn which is liquid at room temperatures. 
The physical properties of GaInSn at 25$^{\circ}$C 
are: density $\rho=6.36 \times 10^3$\,kg/m$^3$, kinematic viscosity 
$\nu=3.40\times 10^{-7}$\,m$^2$/s, 
electrical conductivity $\sigma=3.27\times 10^6$\,($\Omega$ m)$^{-1}$. 
At the top and bottom, the liquid metal column is contacted 
by two massive copper electrodes of height 9 cm which are connected 
by water cooled copper tubes to a DC power supply that is able to provide 
electrical currents of up to 8 kA. By intensively rubbing the 
GaInSn into the copper, we have provided a good wetting
making the electrical contact 
as homogeneous as possible.

While in later experiments it is planned to use 
Ultrasonic Doppler Velocimetry (UDV) in order to measure
directly the axial velocity component along the $z$-axis, 
for 
the first experiments we have decided not to use any inserts that 
could possibly disturb the homogeneous current going from the 
copper electrodes to the liquid. 
Therefore, for the identification of TI we 
exclusively rely on 14 external 
fluxgate sensors that measure the vertical 
component $B_z$ of the magnetic 
field. Eleven of 
these sensors are aligned along the vertical axis 
(with a spacing of 7.5\,cm), 
while the remaining three sensors are positioned 
along the azimuth in the upper part, approximately at 15\,cm 
from the top electrode. 
The distance of the sensors from the outer rim of the
liquid metal column is 7.5\,cm. This rather large value, which
is certainly not ideal to identify small wavelength perturbations, 
has been chosen in order to prevent any saturation effects of the fluxgate 
sensors in the comparably strong azimuthal field of the 
axial current.

The main goal of our experiment is to study the influence 
of the electric current through the fluid on the growth rate and 
on the amplitude of the magnetic field perturbations. 
This is done without any insert, as well as 
for two different radii of an inner non-conducting cylinder, 
$r_i$=6\,mm and 12.5\,mm, for which we expect a monotonic increase of 
the critical current.

\begin{figure}[h]
\begin{center}
\epsfxsize=8.2cm\epsfbox{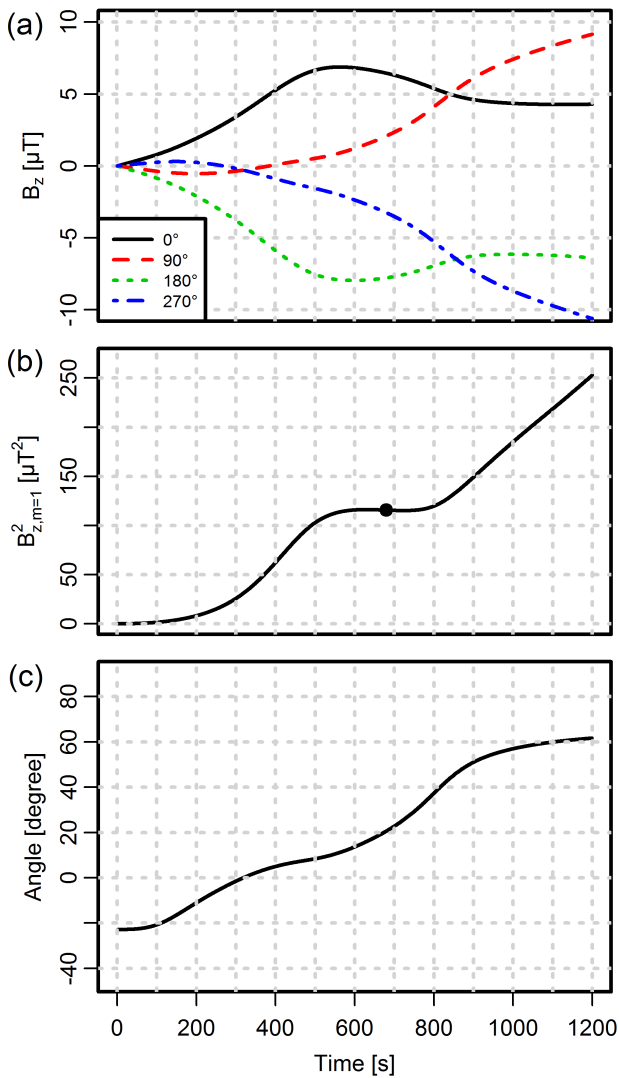}\\
\vspace{2mm}
\caption{Magnetic field data measured along the azimuth,
for $r_i$=12.5 mm and $I$=7 kA.
(a) Time dependence of $B_z$ at 
the 4 fluxgate sensors. The m=1 character
is clearly visible. (b) Squared magnetic field amplitude of 
the m=1 mode. (c) Angle of the m=1 mode.}
\end{center}
\end{figure}

For the particular case with $r_i$=12.5\,mm and $I=7$\,kA, 
Figs.~2 and 3 illustrate the measured $B_z$ data in dependence
on the vertical position and on the azimuth, respectively, and
the procedure of data analysis.
In both cases, the time stamp and the 
field values were set to zero a few seconds after 
the switch-on process of the current had been finalized.
Figure 2a shows the subsequent evolution of $B_z$ measured 
at the 11 vertical positions.
Most of the data in Fig. 2a show a collective long-term trend, 
whose source might be increasing convection, but possibly also
some geometric changes (e.g. by thermal expansion of 
constructional parts) which could influence the projection of the 
strongly dominant azimuthal field component on the 
measured $z$-component. 
 
Apart from this long term trend, the data of Fig.~2a 
contain much more details on the vertical dependence
that can be extracted. For this purpose, we first subtract 
(at every time instant) the mean value and the 
linear trend in $z$-direction and compute then the 
Power Spectral Density (PSD) of the remaining 
signal. The resulting PSD, in dependence on time and 
wavelength, is shown in Fig.~2b.
At the beginning we observe the simultaneous growth
of three modes with short wavelengths, while later 
a long wavelength mode becomes dominant.
The time dependencies of four exemplary modes with wavelengths  
18\,cm, 24\,cm,  36\,cm, and 72\,cm are shown separately 
in Fig.~2c. Evidently, after some initial transient (lasting 
approximately until 150\,s), 
the growth of all short wavelength modes 
acquires an exponential character, 
and we can read off the corresponding growth rates 
in the time interval between 200\,s and 280\,s 
(see the dotted lines in Fig. 2c). 
Note the similarity of  the growth rates 
of the  18\,cm and 36\,cm mode which indicates the merging of 
next-neighbor cells (of the pure TI) as it is known from 
numerical simulations of the combined action of TI and 
(thermal) convection due to internal heating.  

This regular exponential growth of the short wavelength modes 
stops suddenly 
at t=280\,s, when the long wavelength mode (here with
wavelength 72 cm)
becomes dominant. Most interestingly, the growth rate of this 
72\,cm mode 
after the transition becomes quite similar to the growth rate of the 
36\,cm mode shortly before the transition. This may indicate a 
sudden doubling of the
wavelength with increasing 
temperature and convection.  

Now we turn to the discussion of the azimuthal dependence
of the induced $B_z$-perturbations, as measured by the four sensors
around the cylinder.   
The behaviour shown in Fig.\,3a clearly 
indicates a non-axisymmetric (m=1) mode of the field with a 
growing amplitude at the beginning and some rotation and further growth 
at later times. The time evolution of the 
squared amplitude $(B_z(0^{\circ})-B_z(180^{\circ}))^2+
(B_z(90^{\circ})-B_z(270^{\circ}))^2$
of this m=1 mode is 
presented in Fig. 3b, and its corresponding angle in Fig.\,3c.

\begin{figure}[t]
\begin{center}
\epsfxsize=8.0cm\epsfbox{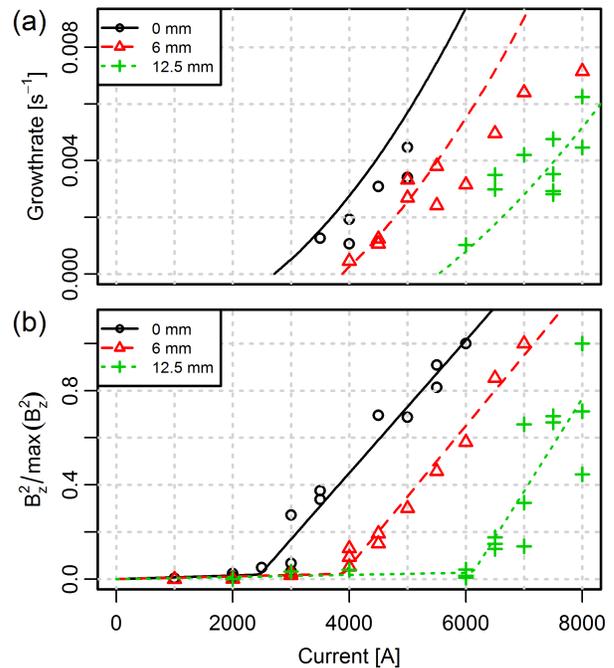}\\
\vspace{2mm}
\caption{(a) Measured growth rates and their numerical 
predictions for three 
different radii of the 
inner cylinder, in dependence on the current. 
(b) Measured squared magnetic field strength of 
the m=1 mode, and their piecewise linear fits. The maximum
squared fields for $r_i$=0\,mm, 6\,mm, and 12.5\,mm 
are 61.0\,$\mu$T$^2$, 543.3\,$\mu$T$^2$, 
and 1668.1\,$\mu$T$^2$, respectively.}
\end{center}
\end{figure}

In Fig.\,4 we compile the dependence of various quantities 
on the radius of the inner cylinder and 
on the current through the fluid. The growth rates of the 
18 cm mode, as extracted 
from the periods with clear exponential growth (see Fig.\,2c 
for one example), are compared with the numerically predicted 
growth rates in Fig.\,4a.  Despite some scatter of the 
data, we observe a quite
reasonable agreement with the numerical predictions.

The second quantity of interest is the saturation level of the 
magnetic field. Similar to many other instabilities,
for TI one would expect a more ore less linear behaviour of the 
saturated squared magnetic field 
above the critical current, according to
$B^2 \sim (I-I_{\rm crit})$. However, the specific problem of our liquid metal 
experiment is that the saturation of the TI does not rely 
on an {\it intrinsic} back-reaction process, but on 
the {\it extrinsic} stabilizing effect of the increasing 
large scale convection. With this caveat in mind, 
in Fig.\,4b we try to identify the critical current 
by plotting the squared magnetic field of the m=1 mode
at the inflection point of the corresponding curves 
(indicated, for this example, by the point in Fig.\,3b). 

Starting at this inflection point we typically observe also a  
drift of the angle of the field pattern. We hypothesize that this
''mode locking''  
starts when the long wavelength mode 
becomes dominant and aligns to some  preferred
azimuthal direction (determined, e.g., by slight 
geometric imperfections of the experimental set-up). The 
observed magnetic field level at this inflection point
is thus yet determined by the TI alone.
Insofar, it is reasonable to assume a monotonic, if not linear,
dependence of the quadratic field strength (taken at the 
inflection point) on $(I-I_{\rm crit})$ to characterize the TI.

\begin{figure}[t]
\begin{center}
\epsfxsize=8.0cm\epsfbox{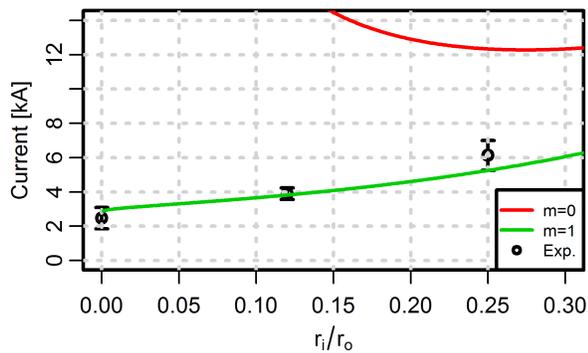}
\vspace{2mm}
\caption{Comparison of the determined critical currents  
with the numerical predictions for the m=1 mode. 
The numerical prediction for the m=0 (sausage) mode 
is indicated partly.}
\end{center}
\end{figure}

By applying a piecewise linear fit to the observed squared 
field intensities (see Fig.\,4b), we can determine the critical current 
at the crossing point of the two straight lines.
Figure\,5 shows the critical currents determined 
that way and the error bars (with a 95 per cent confidence level),
in comparison with the numerically determined ones. Despite the fact
that the inflection-point criterion 
for the determination of the saturated magnetic field strength is 
certainly debatable, we obtain again a reasonable correspondence 
with the predicted critical currents.

The investigation of a current driven instability 
in a column of a liquid metal has revealed 
a rather complex behaviour, comprising (after some initial transient) 
an exponential growth of the anticipated TI mode followed by a 
large-scale convection dominated by Joule heating.
From the sharp bends of the amplitude evolution of the short 
wavelength modes (see Fig.\,2c) it becomes apparent that the 
growing intensity of convection limits the further growth of 
the TI.
This is a drawback of the present experimental 
configuration, since it prevents any investigation of the  
{\it intrinsic} saturation mechanism of TI. The theoretical 
picture for this mechanism 
relies on the on-set of strong turbulence (due to the TI) leading to
a radially dependent turbulent resistivity ($\beta$-effect, see \cite{Gellert_2}) 
which, in turn, would result in a modified radial current 
distribution that is just marginally stable against TI. 
The investigation of this interesting saturation 
mechanism would require 
a significant weakening of the role of Joule heating, 
either by choosing a horizontal orientation of the column
(which, however, introduces an explicit symmetry breaking), 
or by an increase  of the radius, and/or by the replacement 
of GaInSn
with liquid sodium, for which TI is expected to occur already 
around 1 kA. Such a large scale sodium experiment, in which TI  
and MRI can be studied together, is planned for the future. 

This work was supported by DFG 
in frame of SFB 609 and SPP 1488, and by the 
Energy Storage Initiative of the Helmholtz society. 
Fruitful discussions with Rainer Arlt, Rainer Hollerbach and 
Manfred Schultz are gratefully acknowledged. We thank 
Bernd Nowak und Thomas Wondrak for technical support.



\begin{thebibliography}{20}

\bibitem{ZAMM} F. Stefani, A. Gailitis, and G. Gerbeth, ZAMM {\bf 88}, 930 (2008).
\bibitem{DYNAMO} A. Gailitis {\it{et al.}}, Phys. Rev. Lett. {\bf{84}}, 4365 (2000);
R. Stieglitz and U. M\"uller, Phys. Fluids {\bf 13}, 561 (2001);
R. Monchaux et al., Phys. Rev. Lett. {\bf 98}, 044502 (2007).
\bibitem{PROMISE} F. Stefani et al., Phys. Rev. Lett. {\bf 97}, 
184502 (2006); F. Stefani et al., Phys. Rev. E {\bf 80}, 066303 (2009).
\bibitem{SISAN} D.R. Sisan, N. Mujica, W.A. Tillotson, Y.M. Huang, 
 W. Dorland, A.B. Hassam, T.M. Antonsen, and D.P. Lathrop,
 Phys.\ Rev.\ Lett.\ {\bf 93}, 114502 (2004).
\bibitem{PRINCETON} M.D. Nornberg, H. Ji, E. Schartman, A. Roach, and J. Goodman, 
Phys. Rev. Lett. {\bf 104}, 074501 (2010).
\bibitem{VANDAKUROVTAYLER} R.J. Tayler, MNRAS {\bf 161}, 365 (1973);
Yu. V. Vandakurov, Sov. Astron. {\bf 16}, 265 (1972).
\bibitem{BERGERSON} W. L. Bergerson, Phys. Rev. Lett. {\bf 96}, 015004 (2006).
\bibitem{SPRUIT} H.C. Spruit, Astron. Astrophys. {\bf 381}, 923 (2002). 
\bibitem{Gellert_1} M. Gellert, G. R\"udiger, and R.~Hollerbach, MNRAS {\bf 414}, 2696 (2011).
\bibitem{MOLL} R. Moll, H.C. Spruit, M. Obergaulinger, Astron. Astrophys. {\bf 492}, 621 (2008). 
\bibitem{BRADWELL}  R.D. Weaver, S.W. Smith, and N.L. Willmann,
J. Electrochem. Soc. {\bf 109}, 653 (1962); 
D.J. Bradwell, Liquid metal batteries: ambipolar electrolysis and
alkaline earth electroalloying cells, PhD Thesis, MIT, 2011.
\bibitem{CONVERSION} F. Stefani, T. Weier, Th. Gundrum, and G. Gerbeth,
Energy Conv. Manag. {\bf 52}, 2982 (2011).
\bibitem{RUEDIGER2} G. R\"udiger, M. Schultz, and M. Gellert, Astron. Nachr. {\bf 332} 17 (2011).
\bibitem{RUEDIGER3} G. R\"udiger and M. Schultz, Astron. Nachr. {\bf 331},
121-129 (2010).
\bibitem{Gellert_2} M. Gellert, G. R\"udiger, Phys. Rev. E {\bf 80}, 046314 (2009).
\end{thebibliography}
\end{document}